\documentclass{piparticle-final}

\usepackage{amsmath}
\usepackage{amsfonts}
\usepackage{amssymb}
\usepackage{graphicx}

\usepackage{cite} 
\usepackage{epstopdf}

\newcommand{\kB}{k_{\rm B}}

\newcommand{\ue}{\text{e}}

\def\eea{\end{eqnarray}}
\def\bea{\begin{eqnarray}}
\def\ee{\end{equation}}
\def\be{\begin{equation}}

\begin{document}

\volume{7}               
\articlenumber{070008}   
\journalyear{2015}       
\editor{C. A. Condat, G. J. Sibona}   
\received{20 November 2015}     
\accepted{17 April 2015}   
\runningauthor{R. Barreto \itshape{et al.}}  
\doi{070008}         

\title{Thermal transport in a 2D stressed nanostructure with mass gradient}

\author{R. Barreto,\cite{inst1,inst2}\hspace{0.5em}  
        M. F. Carusela,\cite{inst1,inst2}\hspace{0.5em}  
        A. Mancardo Viotti,\cite{inst1}\hspace{0.5em}  
        A. G. Monastra\cite{inst1,inst2}\thanks{E-mail: amonast@ungs.edu.ar}}

\pipabstract{
Inspired by some recent molecular dynamics (MD) simulations and experiments on suspended graphene nanoribbons, we study a simplified model where the atoms are disposed in a rectangular lattice coupled by nearest neighbor interactions which are quadratic in the interatomic distance. The system has a mechanical strain, and the border atoms are coupled to Langevin thermal baths. Atom masses vary linearly in the longitudinal direction, modeling an isotope or doping distribution. This asymmetry and tension modify thermal properties. Although the atomic interaction is quadratic, the potential is anharmonic in the coordinates. By direct MD simulations and solving Fokker--Planck equations at low temperatures, we can better understand the role of anharmonicities in thermal rectification. We observe an increasing thermal current with an increasing applied mechanical tension. The temperatures and thermal currents vary along the transverse direction. This effect can be useful to establish which parts of the system are 
more sensitive to thermal damage. We also study thermal rectification as a function of strain and system size.
}

\maketitle

\blfootnote{
\begin{theaffiliation}{99}
   \institution{inst1} Instituto de Ciencias, Universidad Nacional de Gral. Sarmiento, J. M. Guti\'errez 1150, 1613 Los Polvorines, Argentina.
   \institution{inst2} Consejo Nacional de Investigaciones Cient\'\i ficas y T\'ecnicas, Argentina.
\end{theaffiliation}
}

\section{Introduction and motivation}

Efficient energy consumption is one of the biggest challenges for modern societies. Moreover, miniaturization of electronic devices together with their increasing computing capabilities make heat production per surface/volume unit tend to increase constantly. New technologies are necessary for efficient heat transport. At nanometric scales and for low-dimensional systems, Fourier´s law is not fullfilled and new phenomena arise. Among them, thermal rectification would make possible to build thermal diodes, being low dimensional systems the best candidates (e.g., atomic chains, graphene) \cite{roberts,dhar,li}.
In particular, graphene nanoribbons (GNR) are promising candidates in nanoelectronics. However, at the chip-level integrated circuits, the power density highly increased, making the thermal managment vital to ensure a stable operation of any practical graphene-based device \cite{zli,pop}

Thermal conductance can be modified by defects, impurities, shapes, geometries, mechanical strains, asymmetries, etc. 
It is known that it is necessary to have an asymmetry on the system to achieve heat rectification \cite{das}. Thus, systems with mass gradients due to dopping concentrations, deposition of heavy molecules or a variable width  are candidates to present this phenomena \cite{roberts, jhu, pereira,chang}.

On the other hand, thermal conductance of 2D systems, as GNRs, is remarkably affected by tensile strain. Moreover, mechanical tension is relatively easy to control experimentally at nanoscale, being a good candidate to tune the phononic heat transport in a system \cite{wei,guna,Kgrafeno,Kim}.

From this motivation, we present a simple model for a 2D system to better understand heat transport and the possibility of thermal rectification in a layered-device with variable widths subject to a mechanical longitudinal tension.

\section{The model}

The system is composed of a rectangular array of $N = N_x \times N_y$ particles. They are identified by an index $\bold{i} = (i_x, i_y)$. We consider a linear mass gradient along the $x$ direction, with $m_{\mathbf{i}} = M_L - (i_x -1) (M_L - M_R)/(N_x - 1)$, being $M_L$ and $M_R$ the masses of the left and right columns respectively (see Fig. \ref{figModel}). The particles interact isotropically to nearest neighbors by potentials that only depend on their relative distance $r = | \bold{R}_{\bold{i}} - \bold{R}_{\bold{j}} |$, as:

\begin{figure}[ht]
\begin{center}
\includegraphics[width=0.98\columnwidth]{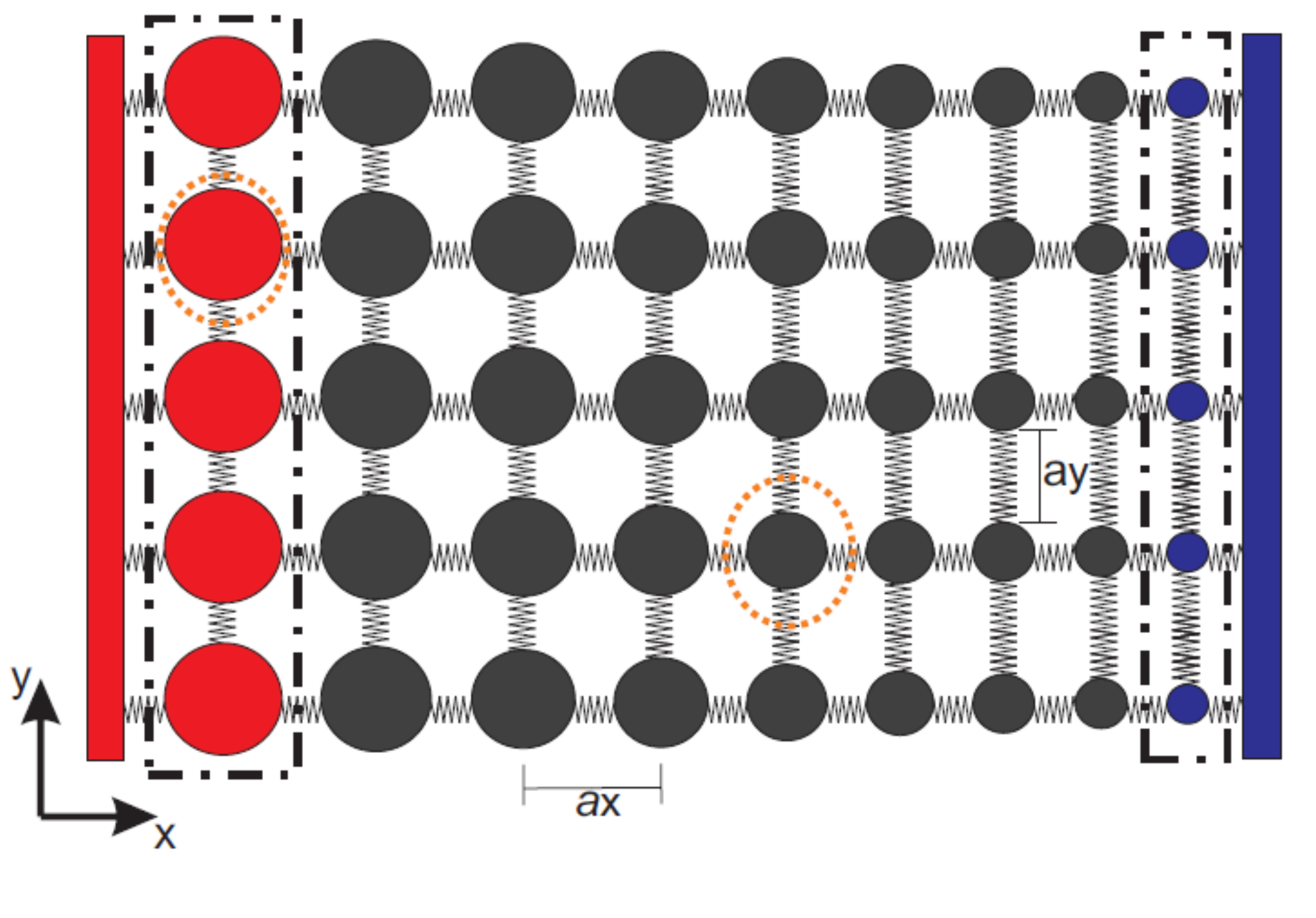}
\end{center}
\caption{Schematic of the system. Particles bordered by dash-dot lines are coupled to Langevin thermal baths.}
\label{figModel}
\end{figure}

\begin{equation} \label{PotX}
v (r) = \frac{1}{2} k (r-l_{0} )^2
\end{equation}

The particles on the left $(i_x = 1)$ and right $(i_x = N_x)$ borders also interact by the same potential with two substracts that can be thought as a left $(i_x = 0)$ and right $(i_x = N_x + 1)$ columns of fixed particles. Therefore, there are $(N_x +1) N_y$ interactions or bonds in the $x$-direction, and $N_x (N_y -1)$ bonds in the $y$-direction that contribute to the total potential 

\begin{align} \label{TotalPot}
V (\{ \bold{R}_{\bold{i}} \} ) &= \sum_{i_x = 0}^{N_x} \sum_{i_y = 1}^{N_y} v (| \bold{R}_{(i_x+1,i_y)} - \bold{R}_{(i_x,i_y)} |) \notag\\
&+ \sum_{i_x = 1}^{N_x} \sum_{i_y = 1}^{N_y -1} v (| \bold{R}_{(i_x,i_y+1)} - \bold{R}_{(i_x,i_y)} |)
\end{align}
that only depends on the positions.

The natural equilibrium position of the particles is a rectangular array with lattice constants $(a_x, a_y) = (l_{0}, l_{0})$. However, if the distance $L_x$ between the fixed rows is bigger than $(N_x + 1) l_{0}$, there will be a tension in the system along the $x$-direction that we parametrize by a change in the lattice constant $a_x > l_{0}$. We are specially interested on the effects of tension on the thermal properties of the system because it is an external parameter that can be easily controlled. The bottom $(i_y = 1)$ and top $(i_y = N_y)$ rows do not interact with any external substrate or reservoir, therefore there could not be any tension in the $y$-direction, being $a_y = l_{0}$ for all cases. The equilibrium position of the particles is $\bold{R}_{0 \bold{i}} = (a_x i_x , a_y (i_y -1))$. We characterize the motion of the particles by the displacement with respect to their equilibrium position $\bold{r}_{\bold{i}} = \bold{R}_{\bold{i}} - \bold{R}_{0 \bold{i}}$. Moreover, the particles on the left 
and right columns are coupled to two thermal reservoirs respectively. We consider a Langevin interaction by a viscous term proportional to velocity, and a decorrelated random force acting on the particles in contact with the reservoirs. The equation of motion for each particle is

\begin{equation} \label{Newton}
m_{\bold{i}} \frac{\text{d}^2 \bold{r}_{\bold{i}} }{\text{d} t^2}  = - \frac{\partial V }{\partial \bold{r}_{\bold{i}} } - \gamma_{\bold{i}} \frac{\text{d} \bold{r}_{\bold{i}} }{\text{d} t} + \bold{f}_{\bold{i}} (t)
\end{equation}
where $\gamma_{\mathbf{i}} = 1$ for $i_x =1$ or $i_x = N_x$, and zero otherwise. The random forces have the correlations $\langle f_{ \mathbf{i}, \mu} (t) f_{ \mathbf{j}, \nu} (t') \rangle = 2 \gamma_{\mathbf{i}} \ \kB T_{\mathbf{i}} \ \delta_{\mathbf{i}, \mathbf{j}} \ \delta_{\mu, \nu} \ \delta (t - t')$, where $T_{\mathbf{i}}$ is $T_{\text{L}}$ and  $T_{\text{R}}$ for $i_x =1$ and $i_x = N_x$, the temperatures of the left and right reservoirs, respectively, or zero otherwise. Indices $\mu$ and $\nu$ run over $x$ and $y$ directions.

For a given realization of the random forces, equations of motion are integrated from some initial condition. After some time, position and velocity of the particles attain a stationary regime where their statistical behavior is constant. 
Averaging a given realization over time, we are mainly interested on the following quantities: i) Mean quadratic velocities: they define a kinetic site temperature $\kB T_{\mathbf{i}} = \frac{1}{2}  m_{\mathbf{i}} ( \langle v_{\mathbf{i},x}^2 \rangle + \langle v_{\mathbf{i},y}^2\rangle)$, even if the system is not in thermodynamic equilibrium; ii) Heat current: for two interacting particles $\mathbf{i}$ and $\mathbf{j}$, energy flows at a rate given by $ J_{\mathbf{i},\mathbf{j}} = \langle F_{\mathbf{i},\mathbf{j}} \cdot ( \mathbf{v}_{\mathbf{i}} +  \mathbf{v}_{\mathbf{j}} ) \rangle $, where $F_{\mathbf{i},\mathbf{j}}$ is the force that $\mathbf{i}$ does on $\mathbf{j}$. The energy current could be different for each bond, being the only conserved quantity the total current through any transversal section of the system; iii) Time correlations: for any component of position or velocity of particle $\mathbf{i}$ over time $ {\cal Q}_{\mathbf{i}} (t)$, we are interested on its autocorrelation $C_{{\cal Q}_{\mathbf{i}}} (\tau) = \langle {\cal Q}_{\mathbf{i}} (t) {\cal Q}_{\mathbf{i}} (t + \tau)  \rangle$. This function contains very useful information about characteristic time scales of the system, how long time averages should be done to be significant, and to estimate the statistical errors of averaged quantities; iv) Spatial correlations $\langle {\cal Q}_{\mathbf{i}} (t) {\cal Q}_{\mathbf{j}} (t)  \rangle$: the motion of particle $\mathbf{i}$ will be in general correlated with the motion of particle $\mathbf{j}$. It is expected that neighbor particles will have stronger correlation than distant particles. A correlation lenght arises, depending on parameters of the system and thermal baths.

\section{Fokker--Planck equations - Harmonic approximation}

Expanding interatomic potential (\ref{PotX}) for two neighboring particles, it has the form $v (r) = \frac{1}{2} k ( r^2 - 2 r l_0 + l_0^2 )$. The term on $r^2$ is fully quadratic on the components of $\bold{r}_{\bold{i}}$ and $\bold{r}_{\bold{j}}$. The problem arises with the second term proportional to $r$, 

\begin{equation} \nonumber
r = | \bold{R}_{\bold{j}} - \bold{R}_{\bold{i}} | = | ( \bold{R}_{0 \bold{j}} - \bold{R}_{0 \bold{i}} ) + (\bold{r}_{\bold{j}} - \bold{r}_{\bold{i}}) | = | \bold{a} + \Delta \bold{r} |
\end{equation}
where, for simplification, $\bold{a}$ is the vector joining the equilibrium positions of the particles, and $\Delta \bold{r}$ the relative displacement between them. We see that this term is not linear on the components of $\bold{r}_{\bold{i}}$ or $\bold{r}_{\bold{j}}$ 

\begin{align} \nonumber
r &= \sqrt{ (\bold{a} + \Delta \bold{r}) \cdot (\bold{a} + \Delta \bold{r})  } \notag\\
&= \sqrt{ a^2 + 2 \ \bold{a} \cdot  \Delta \bold{r} + | \Delta \bold{r} |^2 } \notag\\
&= a \sqrt{ 1 + 2 \ \hat{\bold{a}} \cdot \boldsymbol{\delta} + \boldsymbol{\delta} \cdot \boldsymbol{\delta} }
\end{align}
due to the square root. In the last equality, we have used $a = | \bold{a} |$, $\hat{\bold{a}} = \bold{a} / a $, and $ \boldsymbol{\delta} = \Delta \bold{r} / a$. Taking into account that the proposed model is valid for small displacements, we can make an expansion considering $| \Delta \bold{r} | \ll a$, or equivalently $ | \boldsymbol{\delta} |  \ll 1 $. Collecting terms of the same order in $\delta$, we arrive to

\begin{align} \nonumber
r = a &\left\{ 1 + \hat{\bold{a}} \cdot \boldsymbol{\delta} + \frac{1}{2} \left[ \delta^2 - ( \hat{\bold{a}} \cdot \boldsymbol{\delta} )^2 \right] \right. \notag\\
&\left. + \frac{1}{2} \left[ ( \hat{\bold{a}} \cdot \boldsymbol{\delta} )^3 - ( \hat{\bold{a}} \cdot \boldsymbol{\delta} ) \delta^2 \right] + {\cal O} ( \delta^4 )  \right\}
\end{align}
We see explicitly in this expansion that the potential energy is not quadratic on the coordinates. Nevertheless, for very low temperatures, where displacements are small with respect to the lattice constants, we can neglect the cubic and higher order terms. Making this approximation for all bonds in $x$ and $y$ direction, after collecting terms, we arrive to a quadratic approximation for the total potential energy of the system

\begin{align} \label{TotalPotB}
V (\{ \bold{r}_{\bold{i}} \} ) &\approx (N_x + 1) N_y v_0 \notag\\
&+ \sum_{i_x = 1}^{N_x} \sum_{i_y = 1}^{N_y} \left[ k \ x^2_{(i_x,i_y)} + (k_{\perp} + \alpha_{i_y} k ) y^2_{(i_x,i_y)} \right] \notag \\
 & - \sum_{i_x = 1}^{N_x - 1} \sum_{i_y = 1}^{N_y} \left[ k \  x_{(i_x,i_y)} x_{(i_x + 1,i_y)} \right.\notag\\
 & \ \ \ \ \ \ \ \ \ \ \ \ \ \ \ \  \left. + k_{\perp} y_{(i_x,i_y)} y_{(i_x + 1,i_y)} \right]  \notag \\
& -\sum_{i_x = 1}^{N_x} \sum_{i_y = 1}^{N_y - 1} k \ y_{(i_x,i_y)} y_{(i_x,i_y + 1)}  
\end{align}
where $v_0 = \frac{1}{2} k (a_x - l_{0} )^2$, $k_{\perp} = k ( 1 - l_{0}/a_x )$, and $\alpha_{i_y} = 1/2$ for $i_y = 1, N_y$ or 1 otherwhise. In this harmonic approximation, the directions $x$ and $y$ are completely decoupled. The dependence on the tension comes through the {\it transversal} spring constant $k_{\perp}$. 

For every particle, we have two degrees of freedom, so the system has $M = 2 N$ total number of degrees of freedom. If we call this coordinates as $\{ q_n \}$, we can rewrite the equations of motion as

\begin{eqnarray} \label{XX}
\dot{q_n} &=& \frac{1}{m_n} p_n \\
\dot{p_n} &=& -\sum_{m = 1}^{M} K_{n m} q_m  - \frac{\gamma_n}{m_n} p_n + f_n (t) 
\end{eqnarray}
where $p_n = m_n \dot{q}_n$ are the momenta, and 

\begin{equation} \nonumber
K_{n m} = \frac{\partial^2 V }{\partial q_n \partial q_m}
\end{equation} 
are elements of a force matrix. This is a set of stochastic linear equations, the so called Fokker--Planck (FP) equations, that can be exactly integrated for a given realization of the random forces $\{ f_n (t) \}$. We explain in more detail the general solution following \cite{Risken}.

By defining $2 M$-dimensional vectors $\bold{X} = (q_1, \ldots, q_M; p_1, \ldots, p_M)$ and $\bold{F} (t) = (0, \ldots, 0; f_1 (t), \ldots, f_M(t) )$ we can further simplify the notation

\begin{equation} \label{FPmatrix}
\dot{\bold{X}} = - \mathbb{A} \bold{X} + \bold{F} (t)
\end{equation}
The matrix $\mathbb{A}$ has the structure

\begin{equation}
\mathbb{A} = \left( \begin{array}{cc}
\bold{0} & -\mathbb{M}^{-1} \\
\mathbb{K} & \mathbb{G} \end{array} \right)
\end{equation}
with $\mathbb{K}$ the force matrix, and the diagonal matrices $\mathbb{M}_{n m} = \delta_{n m} m_n$ and $\mathbb{G}_{n m} = \delta_{n m} \gamma_n / m_n$. This matrix $\mathbb{A}$ can be diagonalized, obtaining $2 M$ complex eigenvalues $\lambda_i$ that come in complex conjugate pairs. Their real part is always positive. The eigenvectors are also complex, and arranging them as columns, we obtain the diagonalizing unitary matrix $\mathbb{U}$, which fulfills $\mathbb{A U} = \mathbb{U A}'$, where $\mathbb{A}'$ is a diagonal matrix with the eigenvalues as elements. Transforming and defining $\bold{X}' = \mathbb{U}^{-1} \bold{X}$, and $\bold{F}' (t) = \mathbb{U}^{-1} \bold{F} (t)$, the matrix Eq. (\ref{FPmatrix}) transforms to

\begin{equation} \nonumber
\dot{\bold{X}}' = - \mathbb{A}' \bold{X}' + \bold{F}' (t)
\end{equation}
Now the FP equations are decoupled, although each component of $\bold{F}' (t)$ is a linear combination of all components $f_n (t)$. 

\begin{equation} \nonumber
\dot{x}'_i = - \lambda_i x'_i + f'_i (t)
\end{equation}

For a given realization of the random forces, and some initial condition, each equation can be integrated, obtaining

\begin{equation} \nonumber
x'_i (t) = \ue^{-\lambda_i t} x'_i (0)  +   \ue^{-\lambda_i t}  \int_0^t   \ue^{\lambda_i t´}   f'_i (t') \text{d} t'
\end{equation}
Taking the solutions for each $x'_i$, and using that $\bold{X} (t) = \mathbb{U} \bold{X}' (t) $, we finally obtain the solutions for positions $q_n$ and momenta $p_n$. For sufficiently  long times, the term proportional to the initial condition will vanish, provided that

\begin{equation} \nonumber
t \gg \tau_{\rm max} = \frac{1}{ {\rm min} ( {\rm Re} (\lambda_i) ) } \  .
\end{equation}

Now we are interested in the statistical behavior of the system, particularly in the stationary regime. With these solutions, one can compute the mean values doing averages over the ensemble of random forces, which are necessary to estimate time and spatial correlations, currents per bond and site temperatures.

Taking into account the correlations of the random forces that depend mostly on bath temperatures, and after some calculations, the time and spatial correlations of all dynamical variables can be computed. First, we define the diagonal matrix $\mathbb{D}_{k l} = 2 \gamma_k \kB T_k \delta_{k l}$, where $T_k$ corresponds to the temperature for the moment component which are coupled to a thermal bath, or zero otherwise. Then, transforming this matrix as $\mathbb{D}' = \mathbb{U}^{-1} \mathbb{D} \ ( \mathbb{U}^{-1} )^T$, and defining a new matrix

\begin{equation} 
\mathbb{H}_{m n} (\tau) = \frac{ \mathbb{D}'_{m n} \ue^{-\lambda_n \tau } }{ \lambda_m + \lambda_n }
\end{equation}
the general correlation function is

\begin{equation}
\langle x_i (t) x_j (t + \tau)  \rangle = ( \mathbb{U} \mathbb{H} \mathbb{U}^T )_{i j}
\end{equation}

\begin{figure*}[ht]
\begin{center}
\includegraphics[width=0.7\textwidth]{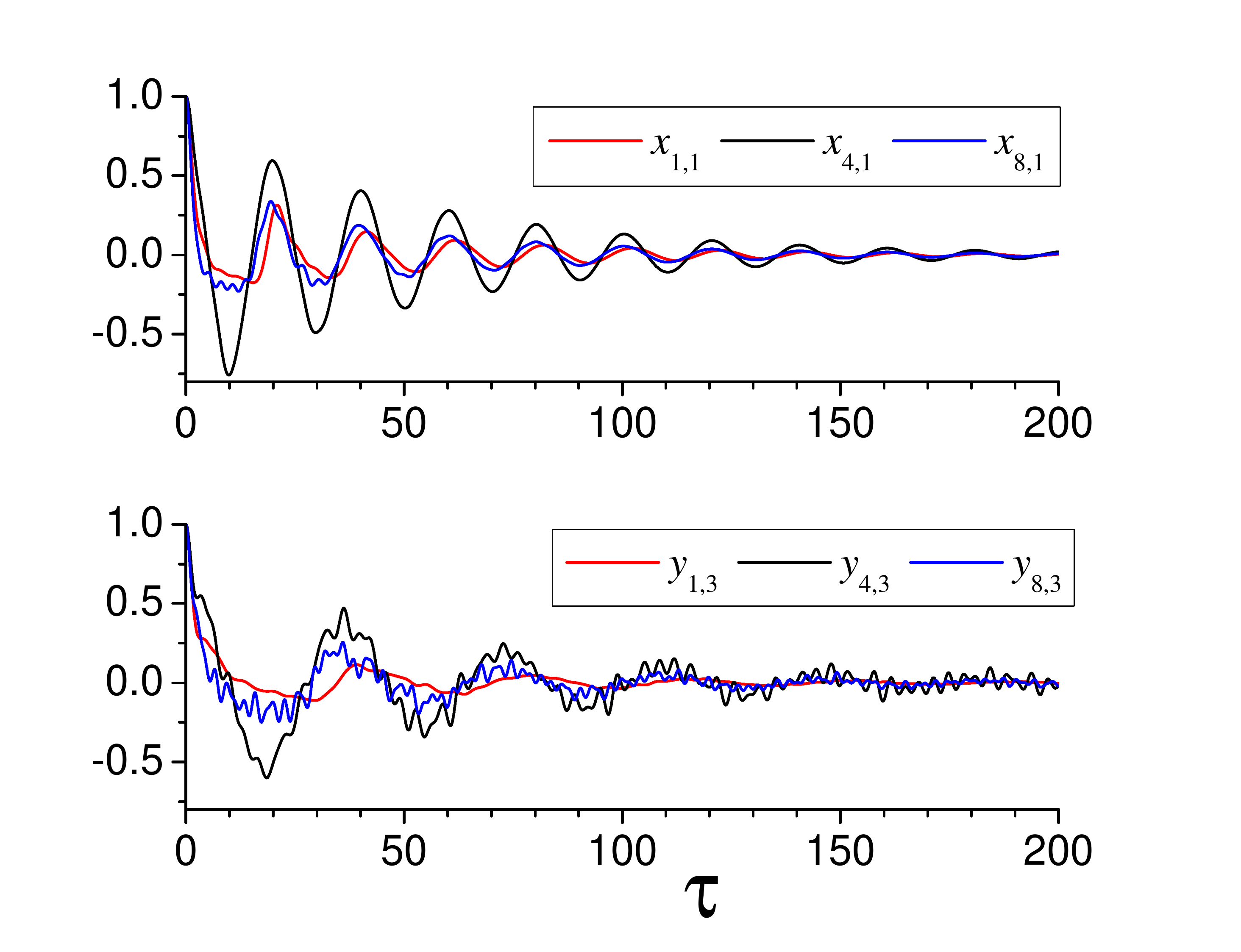}
\end{center}
\caption{Normalized time autocorrelation function for $x$ (up) and $y$ (down) positions of some selected particles. For all figures $M_L = 1.6$, $M_R = 0.4$, spring constant $k =1$, and natural length $ l_{0} = 1$. $(N_x, N_y)=(9,5)$, $T_L = 1.5 \cdot 10^{-4}$, $T_R = 5 \cdot 10^{-5}$, $a_x=1.25$. See final discussion for units.}
\label{figTimeCorrelXFP}
\end{figure*}

\begin{figure*}[ht]
\begin{center}
\includegraphics[width=0.7\textwidth]{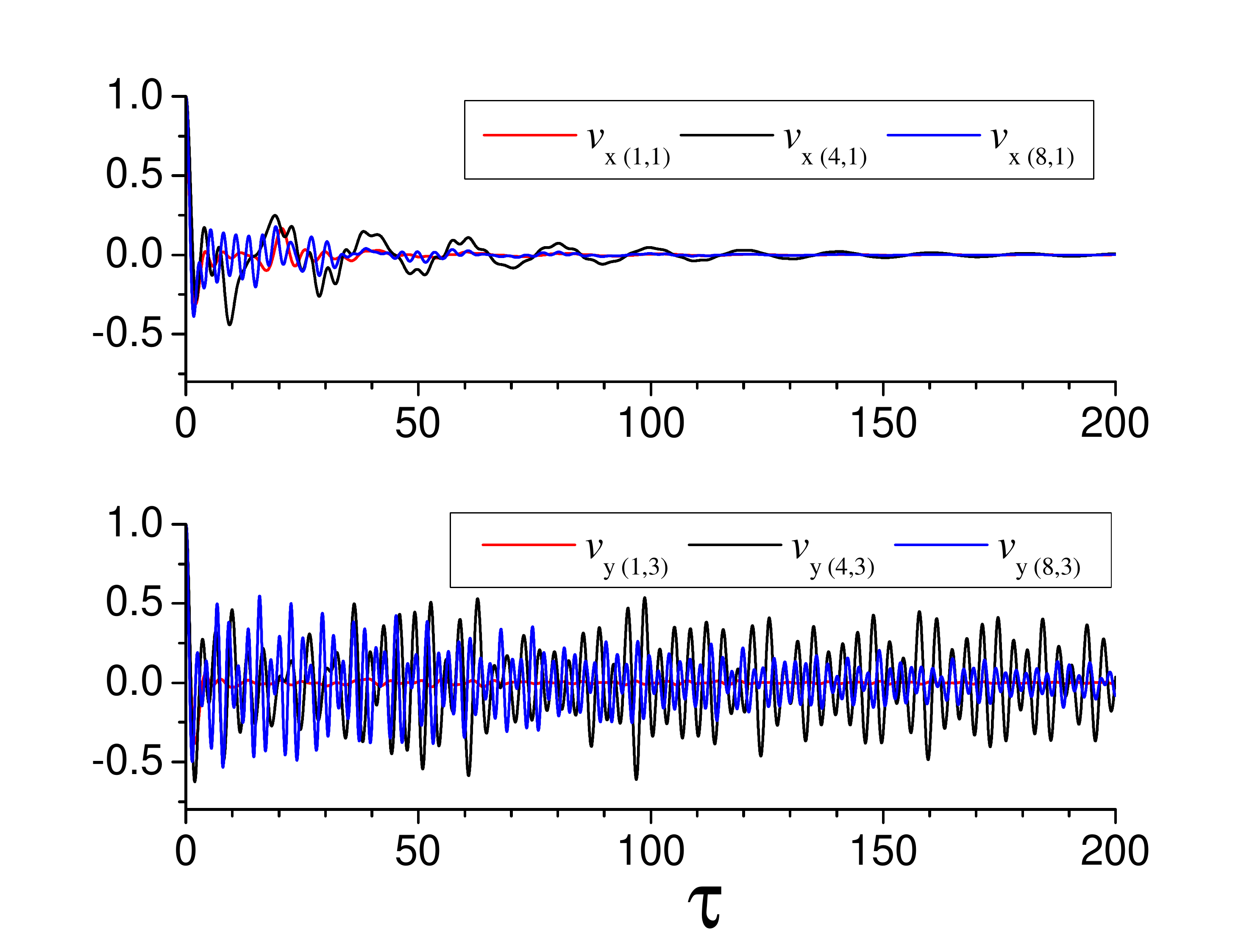}
\end{center}
\caption{Normalized time autocorrelation function for $x$ (up) and $y$ (down) velocities of some selected particles, same values as Fig. \ref{figTimeCorrelXFP}.}
\label{figTimeCorrelVFP}
\end{figure*}

We plot on Figs. \ref{figTimeCorrelXFP} and \ref{figTimeCorrelVFP} the time correlation functions computed by FP equations. We observe that particles coupled to a heat bath, or near to it, decorrelate in shorter times, as expected due to the random forces. Particles in the middle of the system have longer correlation times. The frequencies on these functions are related to normal modes weakly coupled to the heat baths. We observe that for times of the order of 200, most of these functions decay significantly, except for the velocity in $y$ direction. Anyway, the rapid and non periodic fluctuations make it difficult to predict on average the behavior for long times. From these time correlation functions, we conclude that the system can attain a stationary regime at times of the order of 500, and measurements of dynamical variables separated by this time scale can be considered independent.

\section{Numerical results}

The full potential FP equations are difficult to solve. The only way to compute correlations, temperature profiles and heat currents in the stationary state is to perform a direct numerical integration of the equations of motion, as in molecular dynamics (MD) computations, using a Runge-Kutta stochastic integrator. At very low temperatures, where non-linear terms are negligible, FP results coincide with MD within the statistical error. At intermediate temperatures, we first compare the time correlation functions computed by both methods and we observe only minor differences. In general, the functions decay faster for MD, meaning that non-linear terms induce more decorrelation. Nevertheless, we can take the correlation times from FP results as a good upper estimation for all temperature regimes.

\begin{figure}[ht]
\begin{center}
\minipage{0.98\columnwidth}
  \includegraphics[width=0.95\textwidth]{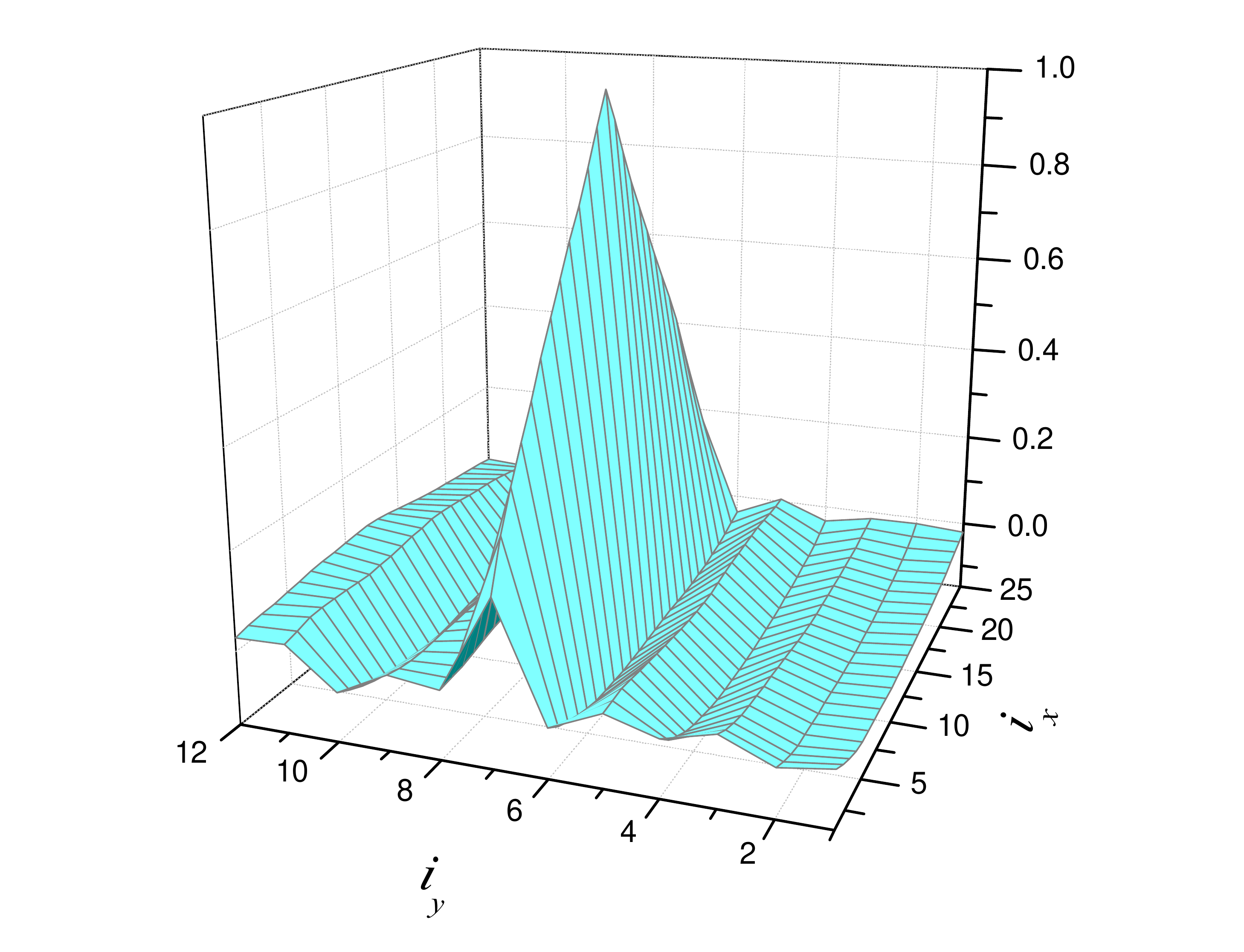}
\endminipage\hfill
\minipage{0.48\textwidth}%
  \includegraphics[width=0.95\textwidth]{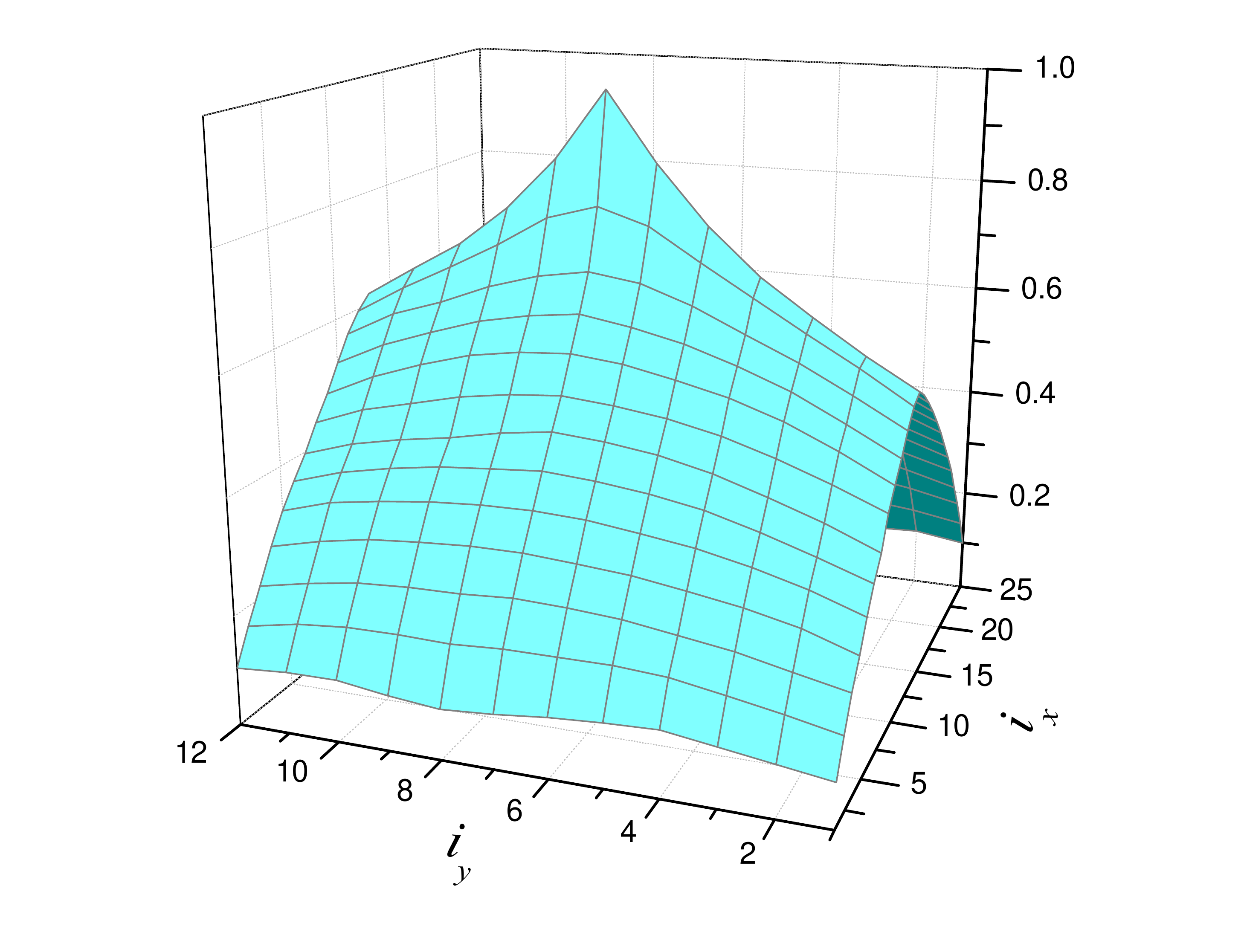}
\endminipage
\end{center}
\caption{Normalized spatial correlation function for $x$ (left) and $y$ (right) position of the $(i_x, i_y) = (12,6)$ particle. $(N_x, N_y)=(25,12)$, $T_L = 1.5 \cdot 10^{-4}$, $T_R = 5 \cdot 10^{-5}$, $a_x=1.25$.}
\label{figSpaceCorrel}
\end{figure}

Spatial correlation functions for the displacements are shown in Fig. \ref{figSpaceCorrel}. The $x$ position of a given particle is correlated with other particles on the same row, while it is almost decorralated with particles in a different row because the coupling is only through higher order terms in the potential (in FP calculations it is strictly zero). On the other hand, the $y$ position of a given particle is strongly correlated with all other particles. Therefore, the correlation length is of the order of the system size for the studied parameters. The spatial correlation of velocities in both directions strongly decays even for two neighboring particles.

\begin{figure}[ht]
\begin{center}
\includegraphics[width=0.98\columnwidth]{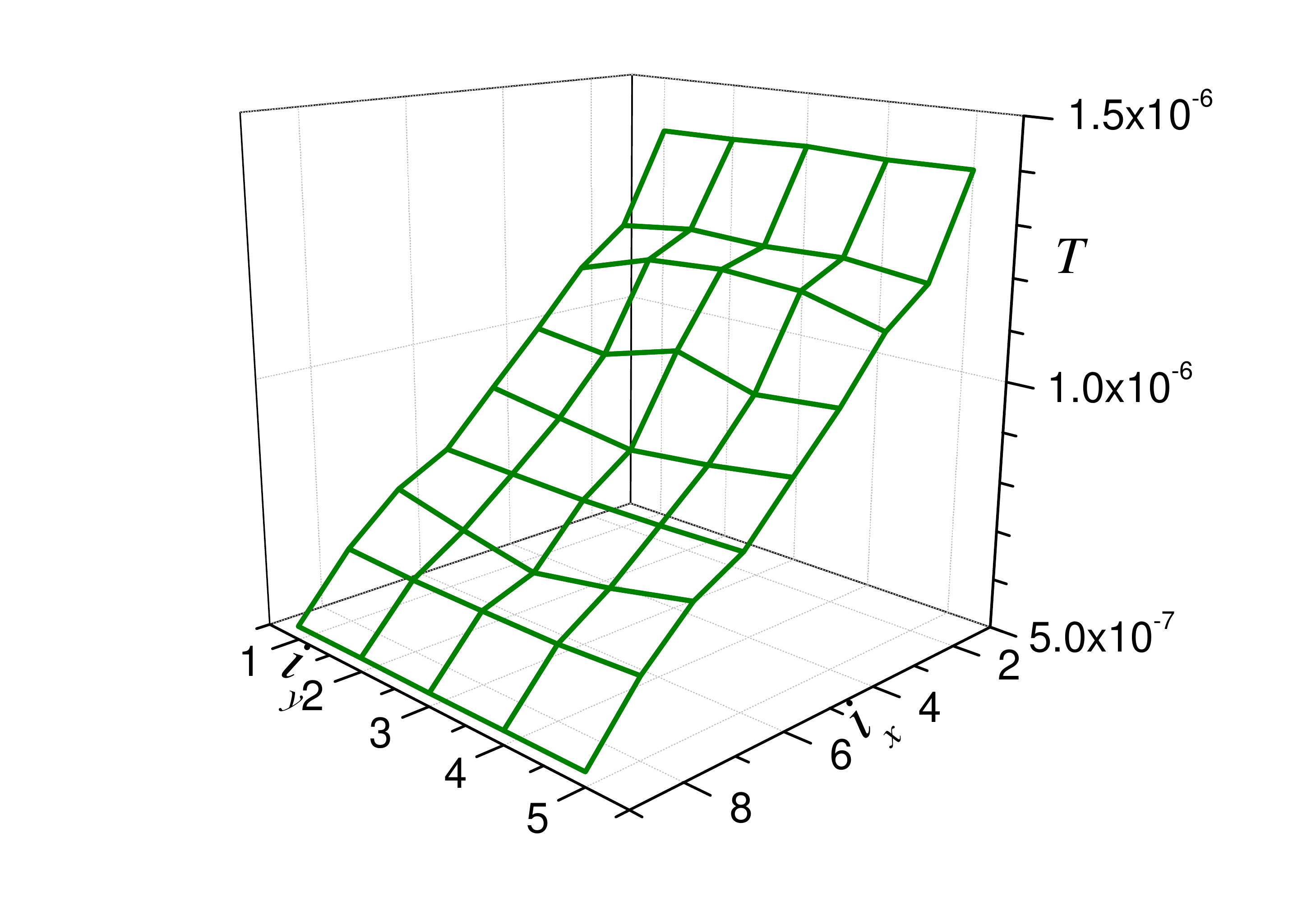}
\end{center}
\caption{Kinetic temperature profile. $(N_x, N_y)=(9,5)$, $T_L = 1.5 \cdot 10^{-6}$, $T_R = 5 \cdot 10^{-7}$, $a_x= 1.2$.}
\label{figTempProfile}
\end{figure}

Averaging the kinetic energy of each particle, we compute temperature profiles, seeing an example in Fig. \ref{figTempProfile}. There is an expected temperature decay in $x$ direction from the hot to the cold bath, although it is not uniform. There are also some fluctuations in the transversal $y$ direction, even at very low temperatures. This is produced by the asymmetry of particles on the top and bottom rows, which are coupled to 3 neighbors, compared to particles on the bulk rows with 4 neighbors.

\begin{figure}[ht]
\begin{center}
\includegraphics[width=0.98\columnwidth]{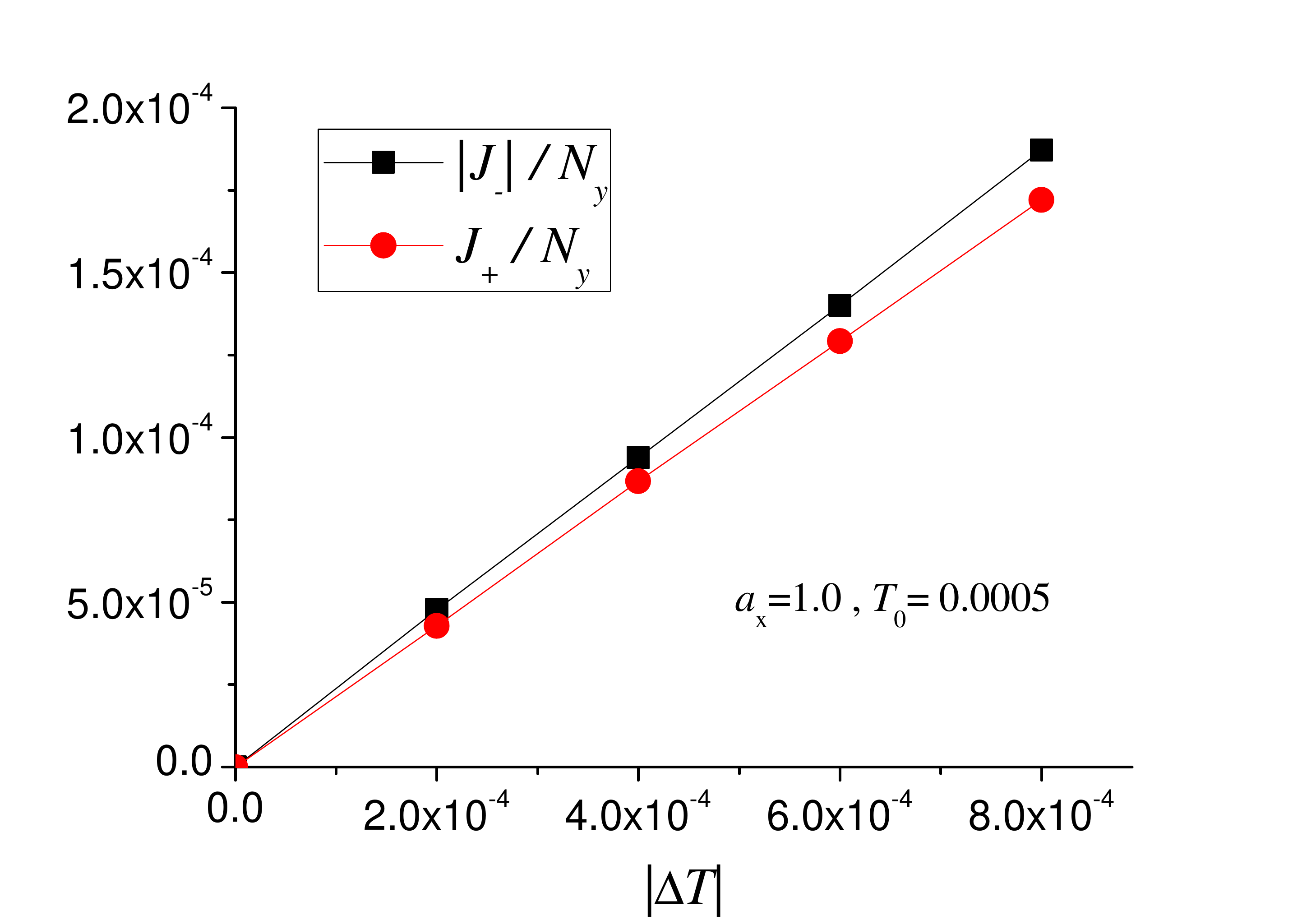}
\end{center}
\caption{Total current per row $J/ N_y$ as a function of  $\Delta T$. $(N_x, N_y)=(25,12)$, $T_0 = 5 \cdot 10^{-4}$,  $a_x = 1.0$.}
\label{figJvsDT}
\end{figure}

The heat currents are computed for every bond. For bonds in $y$ direction, average currents are of the same order of statistical errors, and we can induce that they are almost zero, which is the result given by FP equations. On $x$ direction, the average current is the same for every bond in a given $i_y$ row. On the other hand, the currents on each row are slightly different, depending on strain and other parameters. We finally compute an average heat current per row through the system summing up the currents of all rows $J$, and dividing by $N_y$. We show on Fig. \ref{figJvsDT} this quantity as a function of the temperature bias $\Delta T = T_L - T_R$, for a fixed value of an average bath temperature $T_0 = (T_L + T_R)/2$. We show both results for positive and negative bias, which produce positive and negative heat currents. In both cases, the behavior is linear on $\Delta T$ (allowing us to later define a conductance). However, slopes are different depending the sign of $\Delta T$. This comes from the mass 
gradient that stablishes an asymmetry on the system, and by the non-linear part of the potential. This effect is not observed in the harmonic approximation used to solve FP equations. The asymmetry in the heat transmition is called thermal rectification, and it could be a very useful property.

\begin{figure}[ht]
\begin{center}
\includegraphics[width=0.98\columnwidth]{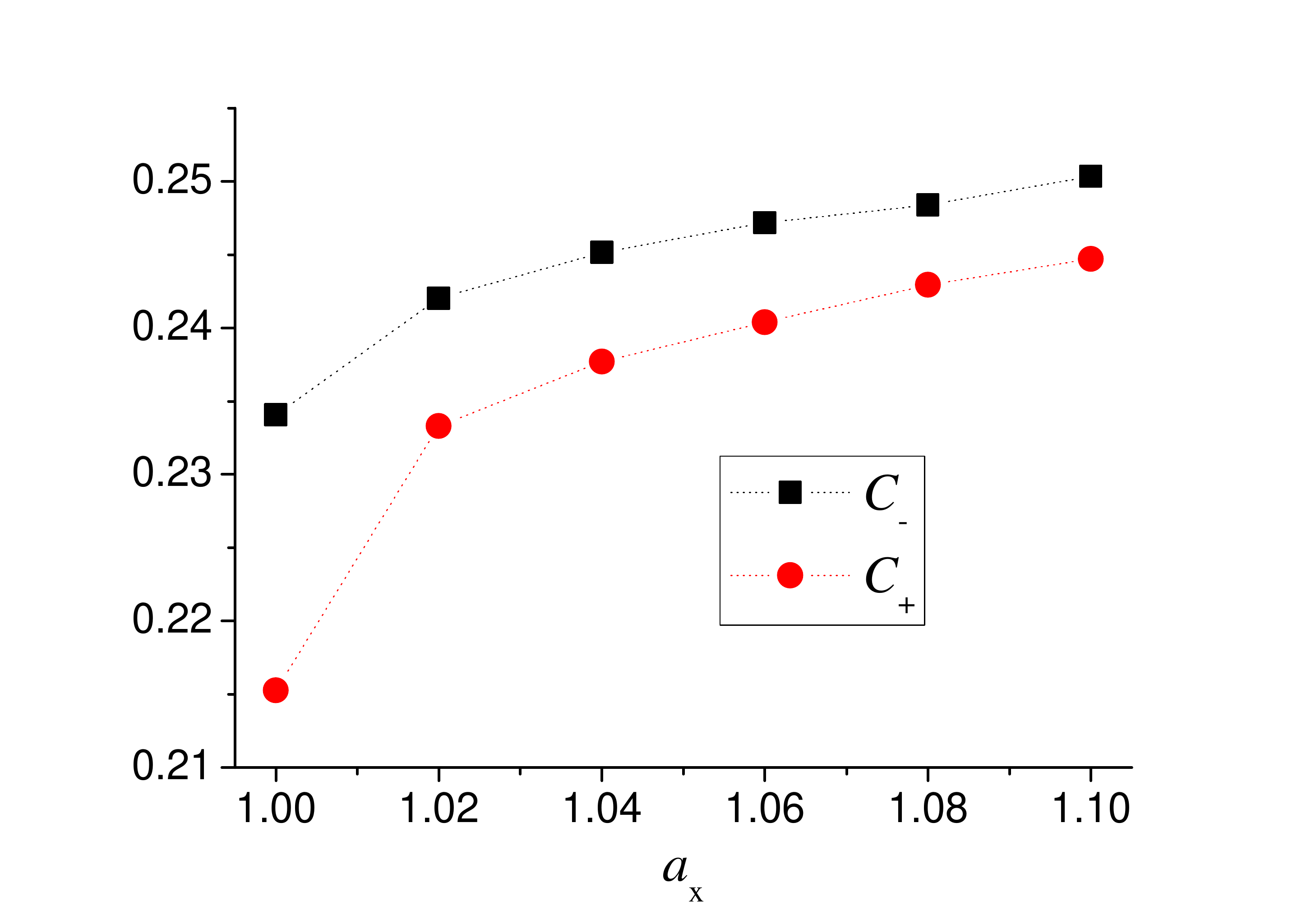}
\end{center}
\caption{Average conductance per row as a function of strain. $(N_x, N_y)=(25,12)$, $T_0 = 5 \cdot 10^{-4}$.}
\label{figCondvsStrain}
\end{figure}

To study the heat current and thermal rectification as a function of strain $a_x$, we plot in Fig. \ref{figCondvsStrain}  the conductance per row ${\cal C} = J / (N_y \Delta T)$, for both signs of the temperature bias. We observe an increasing current as strain is increased, although the thermal rectification seems to decrease as both curves approach.  

\begin{figure}[ht]
\begin{center}
\includegraphics[width=0.98\columnwidth]{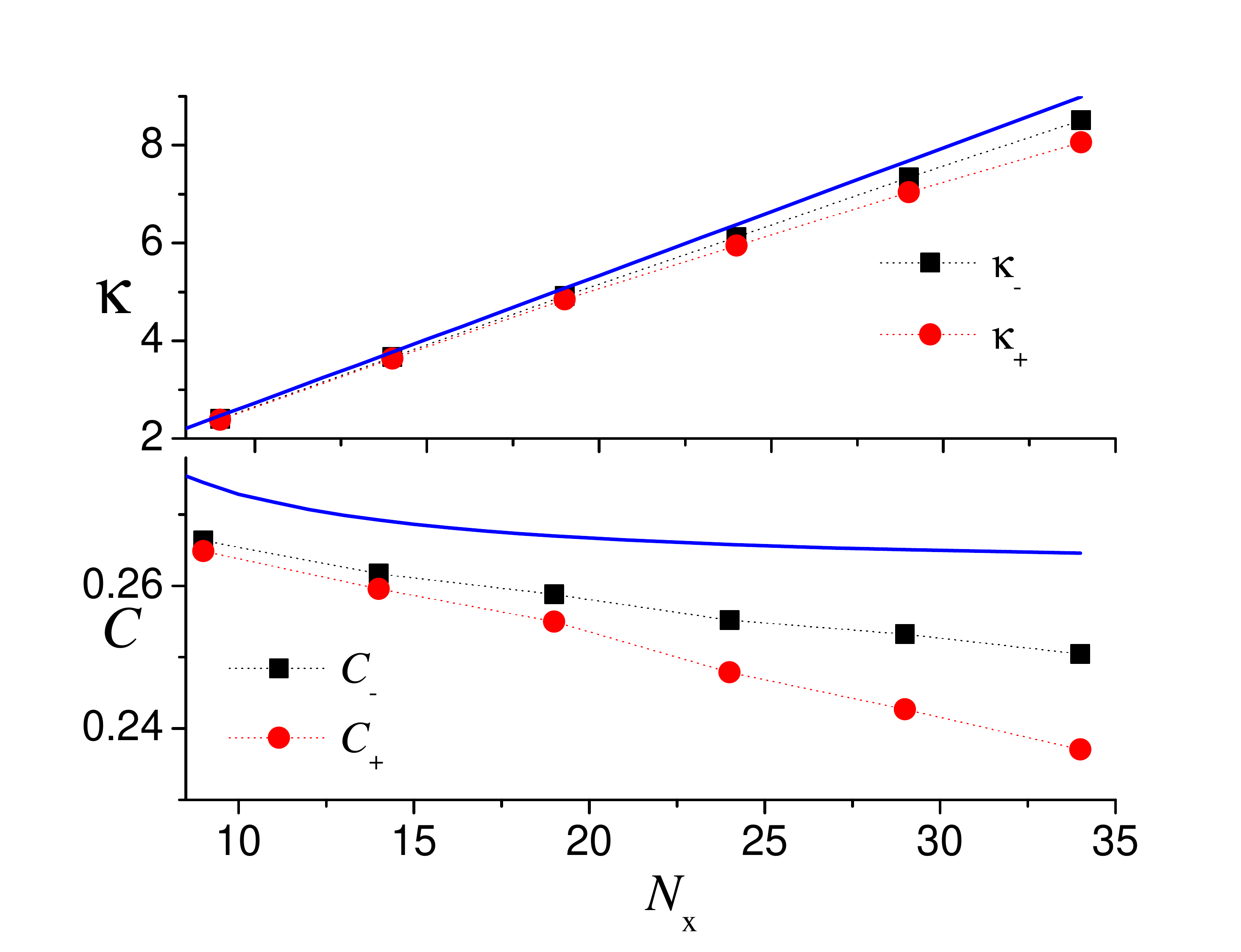}
\end{center}
\caption{Average conductivity (top) and conductance (bottom) per chain as a function of system length. Full (blue) line corresponds to Fokker--Planck calculation. $N_y=5$,  $T_0 = 5 \cdot 10^{-4}$, $a_x = 1.0$.}
\label{figCondvsSize}
\end{figure}

As a funtion of system width $N_y$, the conductance quickly achieves a constant value, being compatible with Fourier law, where heat current is proportional to transversal area. In Fig. \ref{figCondvsSize}, we plot the conductivity and conductance per row as a function of system length $N_x$. FP calculations give a decreasing conductance although it seems to achieve a constant value, breaking the classical Fourier law. Indeed, the conductivity diverges as it is expected for ordered harmonic crystals. For MD computations, we observe a smaller conductance with respect to FP, and it decreases faster with system size. We observe that thermal rectification increases with length, at least for this small system regime, which can be a very useful effect to build a thermal diode. Whether or not these tendencies continue in the thermodynamic limit ($N_x \rightarrow \infty$) cannot be inferred from these simulations; and answering this question is beyond the scope of the present work. 

\section{Conclusion and discussion}

We studied thermal properties of a 2D atomic model with mass gradient under tension. Although it is a simple theoretical model, we can make some relations with graphene, a real 2D atomic system, at least for some orders of magnitude. For thermal and mechanical properties, graphene models use the effective classical Tersoff-Brenner potential. The interaction along carbon-carbon direction can be approximated by a quadratic potential with constant $k_r$ = 652 nN/nm and equilibrium distance $a_0$=0.1421 nm \cite{Kgrafeno}. Taking into account the carbon mass $m_0$, we obtain time and temperature scales $\tau_0=\sqrt{m_0/k_r}$= 5.5 fs , and $T_0 = k_r a_0^2 / \kB \approx 10^6$ K.

Autocorrelation functions in our model would suggest a time scale from 3 to 6 ps for loss of information, at least for systems of length up to 30 atoms. Characteristic frequencies can be related to those of normal modes weakly coupled to thermal baths.

Room temperature of around 300 K corresponds to dimensionless parameter $T \approx 3 \cdot 10^{-4}$. From some experimental works \cite{Kim}, it is known that suspended graphene could be stable up to 2600 K. In our model, this temperature corresponds to $T \approx 2.7 \cdot 10^{-3}$, where the mean particle displacement becomes of the order of lattice constant. In this case, quadratic approximation for the atomic interaction and nearest neighbor approximation would not be valid anymore. A typical thermal current per atomic chain, as in Fig. \ref{figJvsDT}, would correspond to $2.5 \cdot 10^{-3}$ W. We observed an increasing conductance with uniaxial strain, contrary to MD computations for AGNR (armchair) graphene\cite{guna}. However, as in our model, an increase of conductance was obtained for ZGNR (zigzag) graphene nanoribbons at small strain and room temperature, due to an increased of the phonon velocity of some modes in the low and high frequencies for small strains. 

We observed thermal rectification in a model with an interatomic potential quadratic on the distance, although as a two dimensional model, the potential is non-linear in the coordinates. Also the mass gradient, which makes the system asymmetric, is essential for the thermal rectification. The effect could be even stronger incorporating cubic and quartic terms in the interatomic potentials (Fermi--Pasta--Ulam models), directional terms (strongly present in carbon-carbon interaction), and flexural modes.

\end{document}